\documentclass[12pt,preprint]{aastex}
\received{2005}

\shorttitle{GCRT~J1745$-$3009}
\shortauthors{Hyman et al.}

\newcommand\src{\protect\objectname[]{GCRT~J1745$-$3009}}
\newcommand{\mjybm}{\mbox{mJy~beam${}^{-1}$}}
\newcommand{\kms}{\mbox{km~s${}^{-1}$}}

\begin{document}
\title{A New Radio Detection of the Bursting Source \src}

\author{Scott D.~Hyman}
\affil{Department of Physics and Engineering, Sweet Briar College,
	Sweet Briar, VA  24595}
\email{shyman@sbc.edu}

\author{T.~Joseph~W.~Lazio}
\affil{Remote Sensing Division, Naval Research Laboratory, Washington, DC
	20375-5351}
\email{Joseph.Lazio@nrl.navy.mil}

\author{Subhashis Roy}
\affil{ASTRON, P.O. Box 2, 7990 AA Dwingeloo, The Netherlands.}
\email{roy@astron.nl}

\author{Paul S.~Ray}
\affil{E.~O.~Hulburt Center for Space Research, Naval Research Laboratory, Washington, DC
	20375-5352}
\email{Paul.Ray@nrl.navy.mil}

\and

\author{Namir E.~Kassim}
\affil{Remote Sensing Division, Naval Research Laboratory, Washington, DC
	20375-5351}
\email{Namir.Kassim@nrl.navy.mil}

\begin{abstract}
\src\ is a transient bursting radio source located in the direction of the
Galactic center, discovered in 330~MHz VLA observations from
2002 September 30--October 1 by \citeauthor{hlkrmy-z05}. We have searched for bursting
activity from \src\ in nearly all
of the available 330~MHz VLA observations of the Galactic center since 1989 as well as in
2003 GMRT observations.  
We report a new radio detection of the source in 330~MHz GMRT data taken on
2003 September 28.  A single $\sim$0.5 Jy burst
was detected,
approximately 3$\times$ weaker than
the five bursts detected in 2002.  Due to the sparse sampling of the 2003 observation, only
the decay portion of a single burst was detected. We present additional evidence indicating
that this burst is an isolated one, but we cannot completely 
rule out additional undetected bursts that may have occured with the same 
$\sim$77 min. periodicity observed in 2002 or with a different periodicity.
Assuming the peak emission was detected, the
decay time of the burst, $\sim$2 min, is consistent with that determined for the 2002 bursts.
Based on the total time for which we have
observations, we estimate that the source has a duty cycle of
roughly~10\%. 

\end{abstract}

\keywords{Galaxy: center --- radio continuum --- stars: variable: other}

\section{Introduction}\label{sec:intro}

Transient radio emission has been detected from many astronomical
sources including flare stars, brown dwarfs, masers, gamma ray bursts,
pulsars, supernovae, neutron star and black-hole X-ray binaries, and
active galactic nuclei.
Efficient \emph{blind} searching for radio transients requires a
telescope for which the product of the field of view~$\Omega$, the
sensitivity or collecting area~$A$, and the ratio of the total
observing time to the time resolution~$T/\delta t$ is ``large.''
Generally, radio telescopes have been able to maximize only two of
these three quantities.  Thus, the majority of radio transients have
been found by either monitoring objects thought to be potential radio
emitters (e.g., flare stars and brown dwarfs) or by followup
observations of objects detected at higher energies (e.g., X-ray
binaries and gamma-ray bursts).  Recent developments in low frequency
imaging techniques have produced wide-field images ($\approx
3\arcdeg$~FWHM) with uniform and high resolution across the field
\citep{lklh00,nlkhlbd04} thereby enabling efficient searches for radio
transients \citep{hlkb02,hlknn03}

\begin{figure}
\epsscale{0.45}
\rotatebox{-90}{\plotone{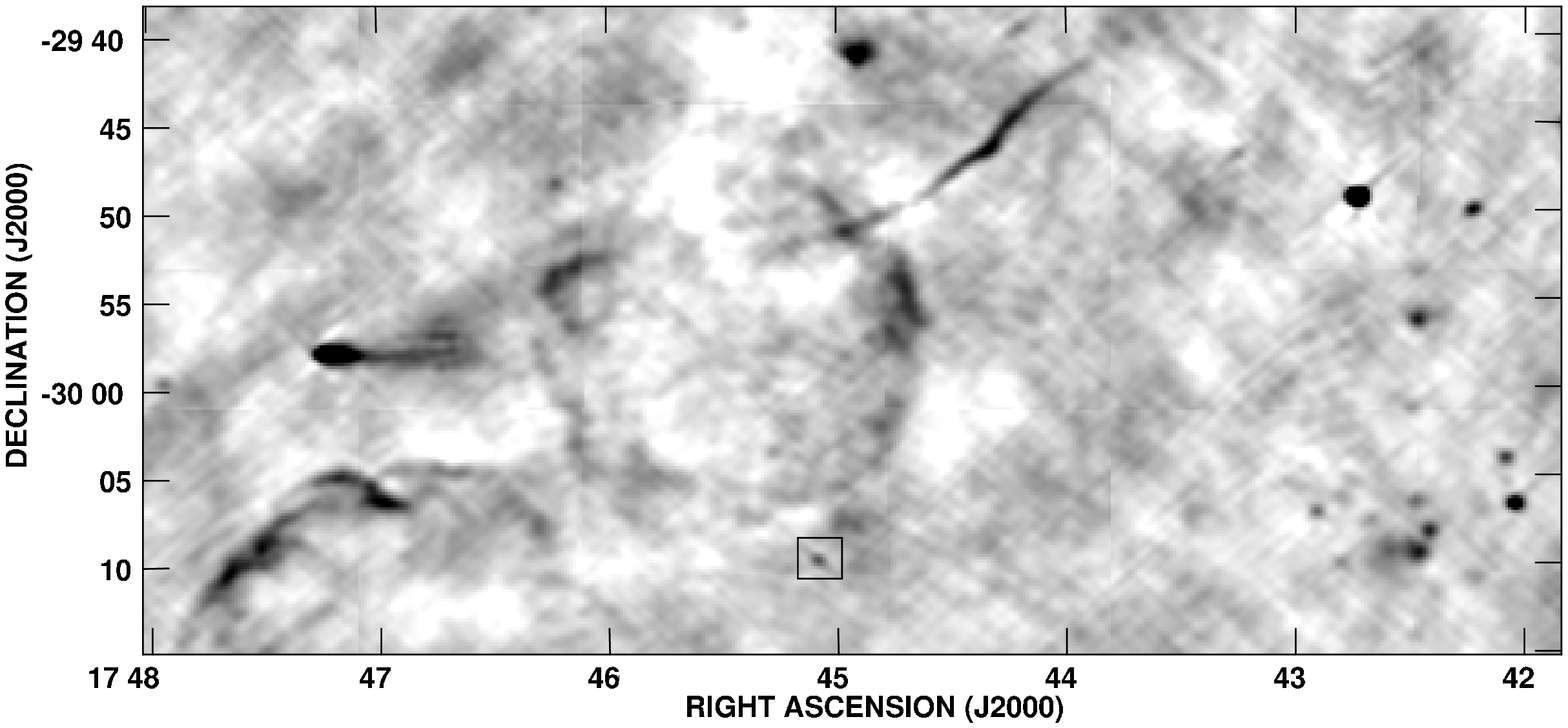}}
\caption[]{Image of the Galactic center field at~330~MHz from the
discovery observations on 2002 September~30 \citep{hlkrmy-z05}.  The
transient source \src\ is indicated by the small box below the
approximately 20\arcmin\ diameter shell of
\protect\objectname[]{SNR~359.1$-$00.5}. The resolution and
sensitivity of the image are $48\arcsec \times 39\arcsec$ and
15~\mjybm, respectively.  \src\ appears as a 100~mJy source here since
it is averaged over five, short ($\sim 10$~min.), roughly 1~Jy bursts
out of a total of a 6-hr observation.  Other sources in the field of
view include the sources to the west which are part of
\protect\objectname[]{Sgr~\hbox{E}}, the \protect\objectname[]{Snake}
is the linear feature to the north, and the
\protect\objectname[]{Mouse} is northeast of \src.}
\label{fig:chart}
\end{figure}

\src\ is a novel bursting radio source \citep{hlkrmy-z05}, whose
notable properties include ``flares'' approximately 1~Jy in magnitude
lasting approximately 10~min.\ each and occurring at apparently
regular 77~min.\
intervals.  This object is located about~1.25\arcdeg\ south of the
Galactic center (GC, Figure~\ref{fig:chart}) and was identified
from~330~MHz (90~cm) observations with the Very Large Array (VLA)
on~2002 September~30.

The source \src\ is notable because it is one of a small number of 
\emph{radio-selected} transients.  Moreover, with only a few
exceptions \citep{m02} such as electron cyclotron masers from flare
stars and the planets, plasma emission from solar radio flares, pulsar
radio emission, and molecular-line masers, most radio transients are
incoherent synchrotron emitters.  For an incoherent synchrotron
emitter, the energy density within the source is limited to an
effective brightness temperature of roughly $10^{12}$~K by the inverse
Compton catastrophe \citep{readhead94}.
The properties of \src\ suggest strongly that its brightness
temperature exceeds $10^{12}$~K by a large factor and that it is a
member of a new class of coherent emitters.

The discovery observations of \src\ were based on VLA 330~MHz observations at a
single epoch, from which only a limited amount of information about
the source could be gleaned.  
This paper reports on a second detection of \src, made with the 
Giant Metrewave Radio Telescope (GMRT) in 2003, as well as on a series of
330~MHz nondetections resulting from archival observations and our Galactic
center radio transient monitoring program.  The observations are summarized in
\S\ref{sec:observe} and the results in 
\S\ref{sec:discuss}.  We discuss briefly the environment of the source
in \S\ref{sec:environment}, and 
we present our conclusions in \S\ref{sec:conclude}.

\section{Observations}\label{sec:observe}

Table~\ref{tab:log} summarizes the 330~MHz observations with the two
telescopes.  Most observations consist of a few, long scans with
occasional interruptions for phase calibration (see below).  At a few
epochs, however, the duration of the observations was not obtained in
a single observation, but in multiple, short and widely spaced
scans. About half of the observations had durations shorter than the
77~min.\ burst period observed in 2002.  At both telescopes, both right-
and left-circular polarization were recorded.

The flux density of \src, even at its peak, is far less than the total
flux density contributed by other sources in the field of view.
Thus, the source can be detected only in images.  In turn, because of
the relatively large fields of view and the number of sources within
the field of view, the entire field of view must be imaged.

Production of the images was conducted in a consistent manner from
epoch to epoch.  Calibration of the flux density was by reference
either to~\objectname[3C]{3C~48} or~\objectname[3C]{3C~286}.  Initial
calibration of the visibility phases was obtained by observations of a
nearby VLA or GMRT calibrator, typically \objectname[]{J1714$-$252}.  At~330~MHz, radio frequency
interference (RFI) can be a substantial problem, and, if not excised
from the visibility data, it would limit the dynamic range of the
final image.  We examined the visibility data for \hbox{RFI} and
excised it.

At~330~MHz, neither the VLA nor the GMRT can be assumed to be
coplanar; in order to image the entire field of view, we used a
polyhedral imaging algorithm to compensate for the non-coplanarity of
the arrays \citep{cp92}.  In order to approach thermal noise limits in
the images, several iterations of imaging, deconvolution
(\textsc{clean}ing), and self-calibration were used.  In order to
search for bursts from \src, the \textsc{clean} components of all
other sources in the field were subtracted from the $u$-$v$ data, and
the residual data were then imaged in 10~min.~subsets.  Noise levels
of the 10~min.\ images range from approximately 10~\mjybm\ for the
GMRT and 20~\mjybm\ for the most extended VLA configurations (A and B) to approximately
250~\mjybm\ for the more compact configurations (C and D), which have
both a lower angular resolution and are more susceptible to
\hbox{RFI} and sidelobe confusion.  If a burst was detected, the residual data then were
imaged with a higher time resolution (from~5 to~30~s) in order to
search for structure within the burst.

All but four of the VLA observations listed in Table~\ref{tab:log} are pointed
in the direction of \objectname[]{Sgr~A*}, approximately
1.25\arcdeg\ north of \src.  (Coincidentally, the discovery observations were
pointed nearly directly at the source.)  The GMRT observations are pointed
approximately 0.5\arcdeg\ west. The primary beam attenuation of the VLA
and the GMRT reduces the apparent flux density of the source by a factor
of approximately 2 and 1.5, respectively.
While significant, this level of primary beam attenuation would not be
sufficient to prevent the recovery of the source, provided that the
amplitude of the bursts is approximately 1~Jy.  However, if
the bursts have a range of amplitudes, significantly weaker bursts
($\lesssim 150$~mJy) could have gone undetected in the vast majority of our 
observations.

\section{Results}\label{sec:discuss}

\begin{figure}
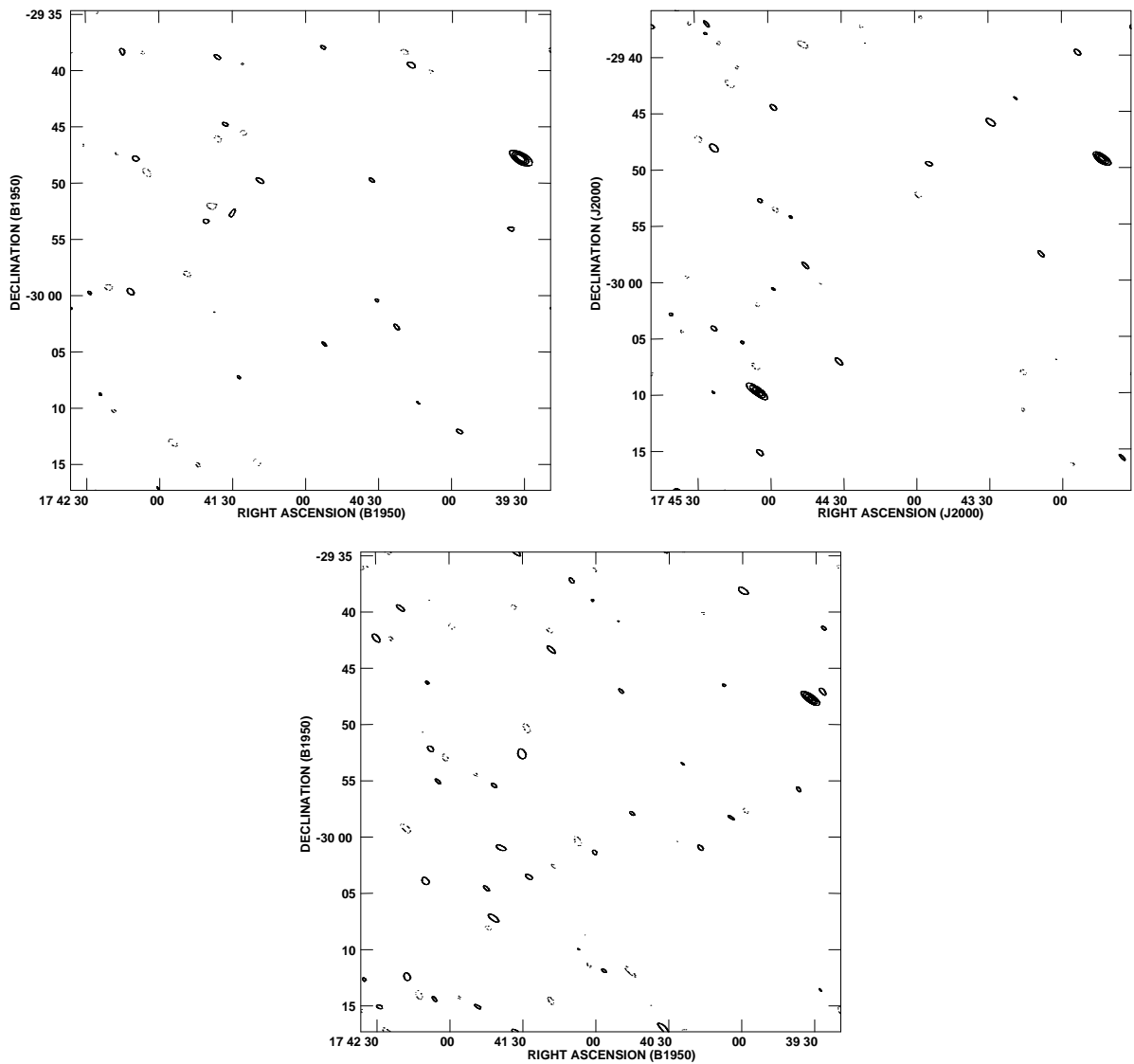

\begin{center}
\epsscale{0.45}
\rotatebox{-90}{\plotone{f2a.ps}}\,
\rotatebox{-90}{\plotone{f2b.ps}}\\
\rotatebox{-90}{\plotone{f2c.ps}}
\end{center}
\caption[]{A 330~MHz VLA image of the field surrounding \src\ and the
source Sgr~E~46 (at upper right in each image) made in 5-min.\
intervals just before (top left), during (top right), and just after (bottom)
the fourth burst detected in~2002 September~30.  The fourth burst is
shown because sampling of it is complete.  Most of the bursts detected
in this epoch were sampled only partially due to (unfortunately-timed)
interruptions for phase calibration.  The transient is located at the
bottom left in the center image. The contour levels are
$-0.4$, 0.4, 0.7, 1.0, and~1.3~Jy~beam${}^{-1}$.}
\label{fig:contour2002}
\end{figure}

\begin{figure}
\epsscale{1.0}
\plotone{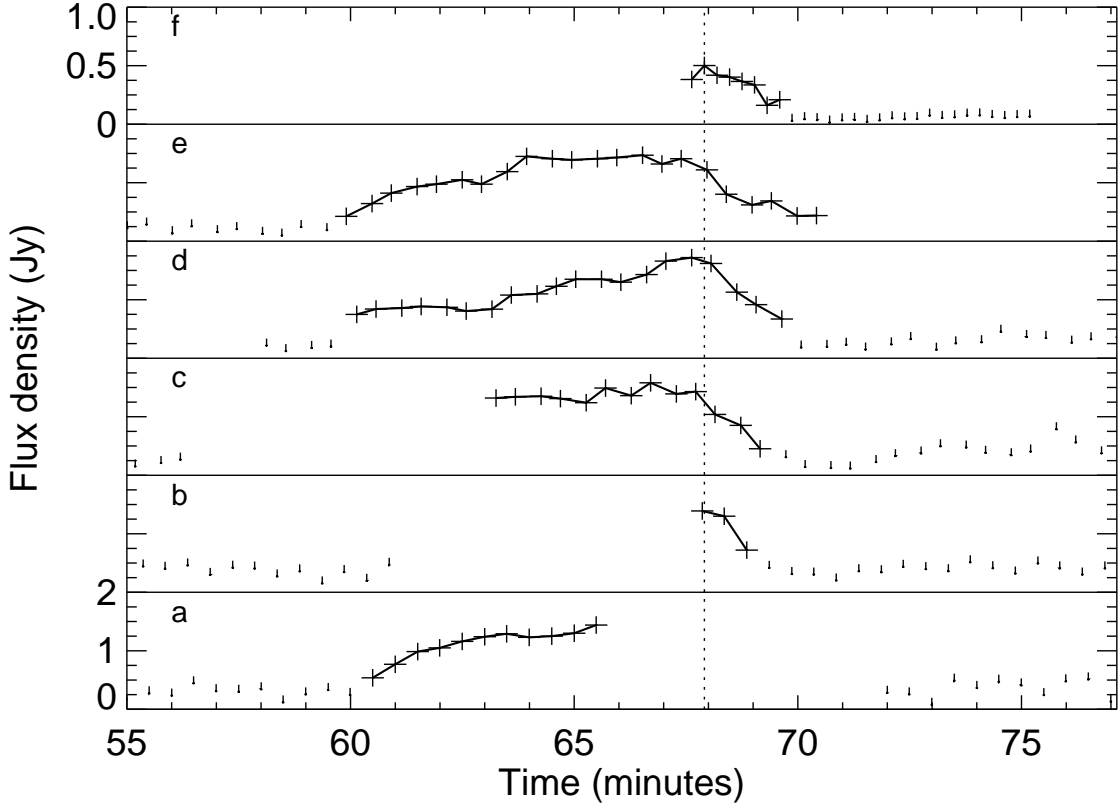}
\caption{The light curves of \src.  The top panel shows the single
detected burst from~2003 September~28 with 17-s sampling; the
remaining panels show the bursts from~2002 September~30 with 30-s
sampling \citep{hlkrmy-z05}, arranged with the fifth burst shown in
the second panel to the first burst in the bottom panel.  For the 2002
September~30 bursts, the light curve has been folded at the apparent
77.1~min.\ periodicity.  For the 2003 September~28 burst, the light
curve has been aligned in time to be consistent with the decay
portions of the 2002 bursts.  In many cases, because the existence of
\src\ was not known at the time of the observation, the full burst is 
not captured because the observations were interrupted for calibration
observations. The 2003 September~28 observations consisted of one 7~min.\ 
scan per hour for several hours. The arrows represent 3$\sigma$ upper limits
for nondetections.} 
\label{fig:lightcurve}
\end{figure}

We detect \src\ at two epochs, 2002 September~30--October~1 and 2003
September~28.  The latter epoch is a new recovery of the source, while
the former epoch is that of the discovery by \cite{hlkrmy-z05}.
Figure~\ref{fig:contour2002} shows contour images before, during, and
after the fourth burst detected on~2002 September~30.
Figure~\ref{fig:lightcurve} shows the light curves for the five 2002
September~30 bursts,
with 30-s sampling, and the 2003 September~28 burst with 17-s sampling.  Unfortunately, the recovery observation
on~2003 September~28 consisted of approximately 10~min.\ scans spaced
approximately an hour apart for several hours.  Only a single burst,
already in its decay phase, is detected at the beginning of a scan
(2003 September~28 11:44:53, IAT).
As shown in Figure~\ref{fig:lightcurve},
the shape of the decay profile for the 2003 September~28 burst is
consistent with that seen for the 2002 September~30 bursts.  Assuming
that this burst is consistent in duration with those from 2002
September~30, the 2003 September~28 burst had a peak of approximately
0.5~Jy, compared to~1--1.5~Jy for those on~2002 September~30.
Clearly, a longer burst duration implies a higher peak flux density.

\begin{figure}
\plotone{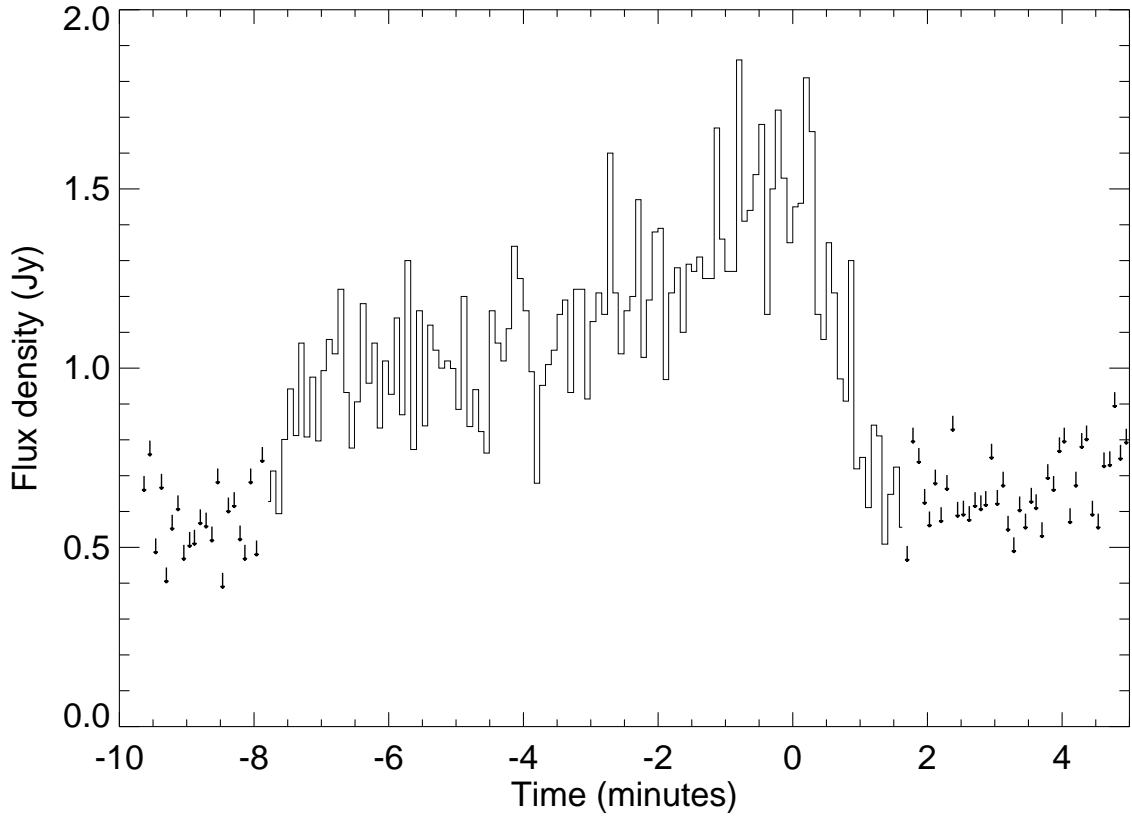}
\caption[]{The fourth burst detected in the 2002 September~30
observations shown with the full 5-s sampling. The arrows represent
3$\sigma$ upper limits for nondetections.}
\label{fig:5seclightcurve}
\end{figure}

Figure~\ref{fig:5seclightcurve} shows the fourth burst from~2002
September~30 with the full 5-s sampling.  The steep decay of the
bursts is depicted much more clearly with higher time resolution.  We
have fitted both the rising and decay portions of the bursts with an
exponential function.  None of the apparent structure in the light
curves (Figure~\ref{fig:lightcurve}) is significant above the
$2\sigma$ level, and no significant structure is evident in the 5-s
light curves that is not also present in the 30-s light curves.

The source is unresolved in both epochs. The angular resolution
($20\arcsec\times10\arcsec$) and sensitivity (50~\mjybm\ for 17-s
integrations) of the 2003 September~28 recovery observation is
significantly improved over the 2002 September~30 discovery
observation.  Fitting a Gaussian to the source yields a position of
(J2000) right ascension $17^{\mathrm{h}}$ $45^{\mathrm{m}}$ 5\fs23 ($\pm
0\fs38$), declination $-30\arcdeg$ 09\arcmin\ 53\arcsec\ ($\pm 5\arcsec$),
which is approximately a factor of two more more accurate in each 
dimension than determined in the 2002 September~30 observation.

Observations at~330~MHz are affected strongly by ionospheric phase
fluctuations.  Their impact includes refractive position shifts.  We
used eight nearby small-diameter sources from the NRAO VLA Sky Survey (NVSS)
\citep{ccgyptb98} to register our images.  The NVSS was conducted
at~1400~MHz and has a substantially better astrometric
accuracy of 0\farcs5 in both right ascension and declination for bright sources.
We found an average ionospheric-induced refraction of $0\fs23 \pm
0\fs24$ in right ascension and $-4\farcs6 \pm 3\farcs4$ in
declination.  The source position and uncertainty cited above include
a correction for this refraction.

No frequency dependence was detected in the 2002 September~30 bursts, and none is detected across the 60-channel, 15-MHz bandpass
for the 2003 September~28 burst.
A power-law fit across the 15-MHz bandpass of the 2003 September~28
observation yields a wavelength dependence of $S \propto \lambda^{4 \pm
5}$.  No circular polarization is detected in the bursts with an upper
limit of 15\% obtained for both epochs.  Linear polarization
measurements are not available for either the discovery or recovery
observations.

No emission is detected from \src\ when imaging the 2003 September~28
observation at times when the burst is not occurring.  We are able to
improve the ($5\sigma$) upper limit for~330~MHz interburst emission
from~75~mJy, for the discovery observations, to~25~mJy, for the
recovery observation.  The upper limit on quiescent emission during
periods of no burst activity is 15~mJy at~330~MHz \citep{hlkrmy-z05}.
Nondetections on~2005 March~25 at both 330 and~1400~MHz also yield an
upper limit of~15~mJy at~330~MHz, but a significantly reduced upper
limit of~0.4~mJy at~1400~MHz, as compared to a 35~mJy upper limit
obtained from a 2003 January observation at that frequency.  We have
also learned that \src\ has been observed in early and mid-2005 with
the Westerbork Synthesis Radio Telescope (WSRT) at both 330
and~1400~MHz.  Upper limits on any emission are approximately a few
milliJanskys (R.~Braun~2005, private communication).

As \cite{hlkrmy-z05} reported, during the discovery epoch (2002
September~30--October~1) the bursts from the source had an approximate
77~min.\ periodicity.  As noted above, the recovery on 2003
September~28 occurs in the midst of a set of~10-min.\ scans, spread
over several hours.  Assuming that the source was emitting periodic
bursts at this epoch, each of 10~min.\ duration, we have determined
the periods at which bursts could occur while being consistent with
the gaps and non-detections during the 2003 September~28 epoch.  There
are 212~min.\ between the detected burst and the end of the
observation.  Thus, we can place no constraints on periods longer than
212~min.  For shorter periods, only the following ranges of periods
are allowed: 36--37~min., 71--78~min., 107--112~min., 131--160~min.,
and 179--195~min.  We estimate that the uncertainty in making these
determinations is perhaps 1~min.\ and results from slightly varying
noise levels within the scans and the assumption that the duration of
the bursts remains fixed at~10~min.  Thus, while consistent with the
gaps and non-detections, the possible 36--37~min.\ periodicity is
perhaps only marginally so. A 77~min.\ periodicity remains consistent
with the 2003 September 28 observations.

We cannot use the interval between~2002 September~30 and~2003
September~28 to constrain the burst activity because the uncertainty
on the 77~min.\ period determined from the 2002 September~30
observations is sufficiently large ($\sim 15$~s) that we cannot
connect the phase between the two observations.  Indeed, the nearest
observation to the 2002 September~30 observation precedes it
by~70~days while the nearest to the 2003 September~28 observation,
other than that on 2003 September~29, follows it by~16~days.  The
current uncertainty is large enough that a single burst could not be
connected in phase over these intervals, even if the source had been
detected.  In addition, the sparse sampling of the observation made
one day after the 2003 September~28 detection does not include any
scan at multiples of 77 min.\ later, nor do the scans, when taken
together with those on September~28, significantly alter the allowed
ranges of other periods given above. Thus, we cannot use the
nondetection on 2003 September~29 to place limits on the duration of
the active period of the source.

The 2003 September~28 and~29 observations were part of a series of GMRT
observations designed to survey the entire Galactic center region.
These include a number of other pointings, not included in Table~\ref{tab:log},
that potentially could be used
to detect \src, albeit with much larger primary beam attenuation.  In
particular, there are scans on \objectname[]{Sgr~A}, some 1\fdg3 away
from both \src\ and the recovery observation's pointing center,
that end approximately 10~min.\ before, 77~min.\ before, and~154~min.\ after the
decay portion of the burst detected in the recovery observation.
We estimated the amount of primary beam
attenuation by measuring the peak flux density of
\emph{\objectname[]{Sgr~A}} from the recovery observation of \src.
The primary beam attenuation is approximately a factor of~10.  Thus,
in the scans on \objectname[]{Sgr~A}, if the transient were bursting
at the level of about~1~Jy, it would appear as a 100~mJy source.  Such
a source would still be well above the noise level in the images for
each scan, but \src\ is not detected in any of them. Thus, the upper
limit for the duration of the detected burst is approximately 13~min.\
and consistent with the 10~min.\ duration observed for the 2002
September~30 bursts. Furthermore, the 77~min.\ period and the range of
other periods allowed by considering only the recovery observation
scans (see above) are largely ruled out by nondetections in the
\objectname[]{Sgr~A} scans from 2003 September 28. Thus, this
burst appears to be an isolated one, in contrast to the 2002
September~30 bursts.

Finally, we note that our northern GC-pointed GMRT scans, centered
$\sim$2\fdg5 to the northeast of \src\ and the bright source, Sgr~E~46
(see Figure~\ref{fig:contour2002}), detect Sgr~E~46, but at a very low level due to severe
primary beam attenuation ($\sim$200$\times$). One of these northern pointings lasted
from 11:36 to 11:44 on 2003 September 28, just before the burst's decay phase
was detected at 11:45 at the beginning of the southern pointing that followed.
Since Sgr~E~46 is located only 0\fdg1
closer to the pointing center than \src, we have corrected the 5~min.\ noise level
at the position of \src\ by the $\sim$200$\times$ primary beam attenuation factor
to crudely estimate the upper
limit of the peak of the 2003 detected burst. An upper limit of $\sim$5~Jy is
obtained, consistent with the 1.5~Jy peak values observed in the 2002 bursts.
However, we consider this result and the evidence that the burst is an 
isolated one to be very tentative, since the location of \src\ is
very far out on the primary beam for these pointings, and since the beam shape
could be asymmetric.

As an initial crude estimate for the duty cycle of the bursting
behavior of \src, we compare the time during which the source is
observed to be active to the total amount of observing time.  The
total observing time is almost exactly 70~hr.  The 2002 September~30
bursts lasted for at least 6~hr; because only a single burst was
detected on~2003 September~28, we assume that the source was active
for~1~hr.  Thus, the apparent duty cycle of \src\ is approximately 10\%.  

Finally, we note that both the original discovery and the recovery
observation occur in late September and are separated by~$\sim$1~yr.
However, given that the discovery and recovery observations occurred
with different telescopes and that there was no detection in our 6-hr 1998
September 25 observation, we can identify no seasonal nor
instrumental explanation that would indicate that the source is not a
celestial object.  By analogy with the model proposed by \cite{tpt05},
it might be the case that the activity of \src\ results from orbital
motion with an approximate 1-yr period.  However, any such model would
also have to explain the non-detections in the 1998 September~25
and~26 epochs.

\section{Environment of \src}\label{sec:environment}

If \src\ is located at the Galactic center, prevailing models explain
it as a compact object, most likely a neutron star
\citep[e.g.,][]{tpt05}.  Motivated by a prediction in
\cite{tpt05}, we have examined images that contain the field
around \src\ in an effort to detect any faint nebulosity, such as
might result from a supernova remnant.  We have examined images
at~330~MHz \citep{lklh00,nlkhlbd04}, 1400~MHz \citep{y-zhc04},
and~2~\micron\ (2MASS); the number of images that we can search is
small because the location of the source is outside of the field of
view of many images of the Galactic center region.

As seen in Figure~\ref{fig:chart}, \src\ is located approximately
10\arcmin\ from the center, and just outside, of the shell-type
supernova remnant \objectname[]{SNR~G359.1$-$0.5} \citep{rf84}.  At a
distance of~$8D_8$~kpc, this angular distance corresponds to a
transverse distance of approximately $25D_8$~pc.  The SNR itself is
old, as evidenced by its size and the extent to which the shell
appears ``broken up.''  Assuming that its age is $10^5T_5$~yr, if
\src\ and the SNR are related, then \src\ would have to have a
velocity of about~$225D_8/T_5$~\kms\ to have reached its current
location.  This velocity is well within those observed for neutron
stars detected as pulsars \citep{acc02}.

In general, there is no diffuse emission surrounding the location
of \src.  One possible exception is some faint emission from the shell
of SNR~G359.1$-$0.5, which lies about~1\arcmin\ north of the position
of the transient.  While this close proximity could be indicative of a
connection between the SNR and \src, there is otherwise no distortion
in the shell of the \hbox{SNR}, akin to that seen for
\objectname[]{G5.4$-$1.2} and \objectname[PSR]{PSR~B1757$-$24}
\citep{fk91}, nor does \src\ have a cometary appearance similar to a
pulsar wind nebula (PWN) like the \objectname[]{Mouse} \citep{gaensleretal04}.

\section{Conclusions}\label{sec:conclude}

We have summarized a series of Very Large Array and
Giant Metrewave Radio Telescope observations of \src\ (Table~\ref{tab:log}).
We detect \src\ at two epochs, 2002 September~30--October~1 and 
2003 September~28.  The latter epoch is a new recovery of the source,
while the former epoch is that of the discovery by \cite{hlkrmy-z05}.

The two sets of detections of \src\ are consistent with the source
producing approximately 1~Jy bursts; we cannot exclude the possibility
that the source produces significantly weaker bursts ($\lesssim
150$~mJy) more frequently. \cite{hlkrmy-z05} reported that the bursts
appear to have a $77.1 \pm 0.3$~min.\ periodicity; we have provided
tentative evidence indicating that the 2003 September~28 burst is an
isolated one.  Given the epochs of observations, we estimate crudely
that the source is active for approximately 7~hr, and the apparent
duty cycle of its activity is roughly 10\%.

We have examined the field around \src\ at radio and infrared wavelengths.
We find possible nebulosity at 1.4~GHz in the shell of
\objectname[]{SNR~G359.1$-$0.5} near the location of the source, but
otherwise no connection between the SNR and the transient.  The
velocity required for \src\ to have originated at the center of
\objectname[]{SNR~G359.1$-$0.5} and reached its current transverse
separation is only roughly 225~\kms.  While well within the range of
velocities observed for various neutron stars, there is also no
compelling reason to think that \src\ and the SNR are related. 

Additional observations are required to determine more about the
nature of \src.  As well as additional searches such as those that we
report here, infrared observations to search for a counterpart, a
periodicity search for weaker pulsed emission, and X-ray observations
to search for quiescent X-ray emission would all be useful.

\acknowledgements

We thank W.~Cotton for providing us with the 20~cm image from
\cite{y-zhc04} to search for possible nebulosity near \src.  This
publication makes use of data products from the Two Micron All Sky
Survey, which is a joint project of the University of Massachusetts
and the Infrared Processing and Analysis Center/California Institute
of Technology, funded by the National Aeronautics and Space
Administration and the National Science Foundation.  The National
Radio Astronomy Observatory is a facility of the National Science
Foundation operated under cooperative agreement by Associated
Universities, Inc.  S.D.H.\ is supported by funding from the Jeffress
Memorial Trust and Research Corporation. Basic research in radio
astronomy at the NRL is supported by the Office of Naval Research.

\clearpage

\clearpage

\begin{deluxetable}{lcccc}
\tablecaption{330 MHz Observational Log\label{tab:log}}
\tablewidth{0pc}
\tabletypesize{\small}
\tablehead{

\multicolumn{2}{c}{Epoch\tablenotemark{a}}
	& \colhead{Telescope\tablenotemark{b}} 
	& \colhead{Bandwidth} 
	& \colhead{Duration} \\

        & 
        & 
	& \colhead{(MHz)}
	& \colhead{(min.)}}
	
\startdata

1989 March~18     & 09:49:40 & VLA:B  &  12.5 & 331.8  \\

\\

1995 October~14   & 22:49:40 & VLA:B  &  1.6 & 30.7 \\
				      
\\

1996 October~19   & 23:03:30 & VLA:A  &  6.2 & 173.5 \\
1996 October~19   & 19:48:50 & VLA:A  &  6.2 & 176.0  \\

\\

1997 February~06  & 13:57:10 & VLA:BnA & 6.2 & 76.7  \\

\\

1998 November~29  & 16:52:20 & VLA:C  &  3.1 & 380.7  \\
1998 September~26 & 02:24:30 & VLA:B  &  3.1 & 124.0  \\
1998 September~25 & 21:08:00 & VLA:B  &  3.1 & 277.5  \\
1998 March~14     & 14:40:50 & VLA:A  &  3.1 & 98.8  \\
1998 March~14     & 10:06:10 & VLA:A  &  3.1 & 231.8  \\

\\

1999 May~31       & 04:53:00 & VLA:D  &  3.1 &  376.7  \\

\\

2001 September~05 & 00:40:60 & VLA:C  &  3.1 & 181.3  \\

\\

2002 March~26     & 10:46:30 & VLA:A  &  6.2 & 29.7  \\
2002 March~26     & 11:18:50 & VLA:A  &  6.2 & 34.7  \\
2002 April~27     & 09:10:40 & VLA:A  &  6.2 & 83.8  \\
2002 May~17       & 08:51:50 & VLA:AB &  6.2 & 83.8  \\
2002 June~24      & 09:22:00 & VLA:B  &  6.2 & 34.5  \\
2002 July~21      & 06:05:60 & VLA:B  &  6.2 & 59.2  \\
2002 September~30\tablenotemark{c}
		  & 20:48:30 & VLA:BC &  6.2 & 289.7  \\
2002 October~01\tablenotemark{c}
		  & 02:34:45 & VLA:BC &  6.2 & 53.8  \\

\\

2003 January~20   & 15:27:40 & VLA:CD &  6.2 & 187.8  \\
2003 July~05      & 08:09:50 & VLA:A  &  6.2 &  34.5  \\
2003 July~08      & 06:28:10 & VLA:A  &  6.2 &  59.2  \\
2003 July~12      & 04:12:50 & VLA:A  &  6.2 &  34.7  \\
2003 July~14      & 04:04:60 & VLA:A  &  6.2 &  34.5  \\
2003 July~28      & 07:09:20 & VLA:A  &  6.2 &  34.5  \\
2003 August~09    & 01:52:50 & VLA:A  &  6.2 &  59.2  \\
2003 August~18    & 23:43:40 & VLA:A  &  6.2 &  59.2  \\
2003 September~28\tablenotemark{c,d}
		  & 10:06:26 & GMRT   &  15 &  63  \\
2003 September~29\tablenotemark{d} 
		  & 11:51:10 & GMRT   &  15 &  45  \\
2003 October~14   & 00:32:50 & VLA:AB &  6.2 &  59.2  \\   
2003 October~21\tablenotemark{d}
	          & 07:27:09 & GMRT   &  15 &  45  \\
2003 October~24   & 00:23:30 & VLA:B  &  6.2 &  59.0  \\
2003 November~23  & 21:16:40 & VLA:B  &  6.2 &  39.7  \\
2003 December~29  & 18:55:10 & VLA:B  &  6.2 &  64.0  \\

\\

2005 March~25     & 13:13:30 & VLA:B &  5.0 & 72.0  \\
			 
\enddata		   
\tablenotetext{a}{We provide the IAT start time of the observation for
use in a later analysis.  However, depending upon the observing
program, the duration of the observation may not have been obtained
in a single observation, but in multiple shorter ones at the epoch.}
\tablenotetext{b}{The notation ``VLA:A'' refers to the A configuration
of the \hbox{VLA}.}
\tablenotetext{c}{Source detected at this epoch.}
\tablenotetext{d}{This epoch consists of multiple, short scans taken over several hours.}
\end{deluxetable}

\end{document}